\documentclass{sig-alternate}

\def \mypapertitle {A Dual Embedding Space Model for Document Ranking}

\usepackage[bookmarks=true%
,bookmarksnumbered=true%
,hypertexnames=false%
,breaklinks=true%
,colorlinks=true%
,linkcolor=blue%
,citecolor=blue%
,urlcolor=blue%
]{hyperref}
\hypersetup{
pdfauthor = {Anonymous},
pdftitle = {\mypapertitle},
pdfsubject = {\mypapertitle},
pdfkeywords = {H.3 [Information Storage and Retrieval]},
pdfcreator = {LaTeX with hyperref package},
pdfproducer = {pdflatex}
}

\usepackage{booktabs} 
\usepackage{graphicx}
\usepackage{url,times}
\usepackage{microtype}
\usepackage{tabularx,amsmath}
\usepackage{amssymb}
\usepackage{multirow}
\usepackage[square,comma,numbers,sort&compress,sectionbib]{natbib}
\usepackage{color}
\usepackage[usenames,dvipsnames]{xcolor}
\usepackage[normalem]{ulem}

 \providecommand{\norm}[1]{\lVert#1\rVert}

\newfont{\mycrnotice}{ptmr8t at 7pt}
 \newfont{\myconfname}{ptmri8t at 7pt}
 \let\crnotice\mycrnotice%
 \let\confname\myconfname%

\toappear{This paper is an extended evaluation and analysis of the model proposed by \citet{nalisnick2016improving} to appear in WWW'16, April 11 - 15, 2016, Montreal, Canada. Copyright 2016 by the author(s).}

\begin{document}


\title{\mypapertitle}

\numberofauthors{4}

\author{
Bhaskar Mitra\\
\affaddr{Microsoft}\\
\affaddr{Cambridge, UK}\\
\email{bmitra@microsoft.com}
\and
Eric Nalisnick\\
\affaddr{University of California}\\
\affaddr{Irvine, USA}\\
\email{enalisni@uci.edu}
\and
Nick Craswell, Rich Caruana\\
\affaddr{Microsoft}\\
\affaddr{Redmond, USA}\\
\email{{nickcr, rcaruana}@microsoft.com}\\
}

\maketitle

\begin{abstract}
A fundamental goal of search engines is to identify, given a query, documents that have relevant text. This is intrinsically difficult because the query and the document may use different vocabulary, or the document may contain query words without being relevant. We investigate neural word embeddings as a source of evidence in document ranking. We train a \emph{word2vec} embedding model on a large unlabelled query corpus, but in contrast to how the model is commonly used, we retain both the \emph{input} and the \emph{output} projections, allowing us to leverage both the embedding spaces to derive richer distributional relationships. During ranking we map the query words into the input space and the document words into the output space, and compute a query-document relevance score by aggregating the cosine similarities across all the query-document word pairs.

We postulate that the proposed \textit{Dual Embedding Space Model} (DESM) captures evidence on whether a document is \emph{about} a query term in addition to what is modelled by traditional term-frequency based approaches. Our experiments show that the DESM can re-rank top documents returned by a commercial Web search engine, like Bing, better than a term-matching based signal like TF-IDF. However, when ranking a larger set of candidate documents, we find the embeddings-based approach is prone to false positives, retrieving documents that are only loosely related to the query. We demonstrate that this problem can be solved effectively by ranking based on a linear mixture of the DESM and the word counting features.
\end{abstract}

\vspace{-0.6\baselineskip}
\vspace{-0.5\baselineskip}
\subsection*{Keywords: \textrm{Document ranking; Word embeddings; Word2vec}}
\vspace{.5\baselineskip}


\section{Introduction}
\label{sec:intro}

\begin{figure}[t]
\center
\includegraphics[width=2.2in]{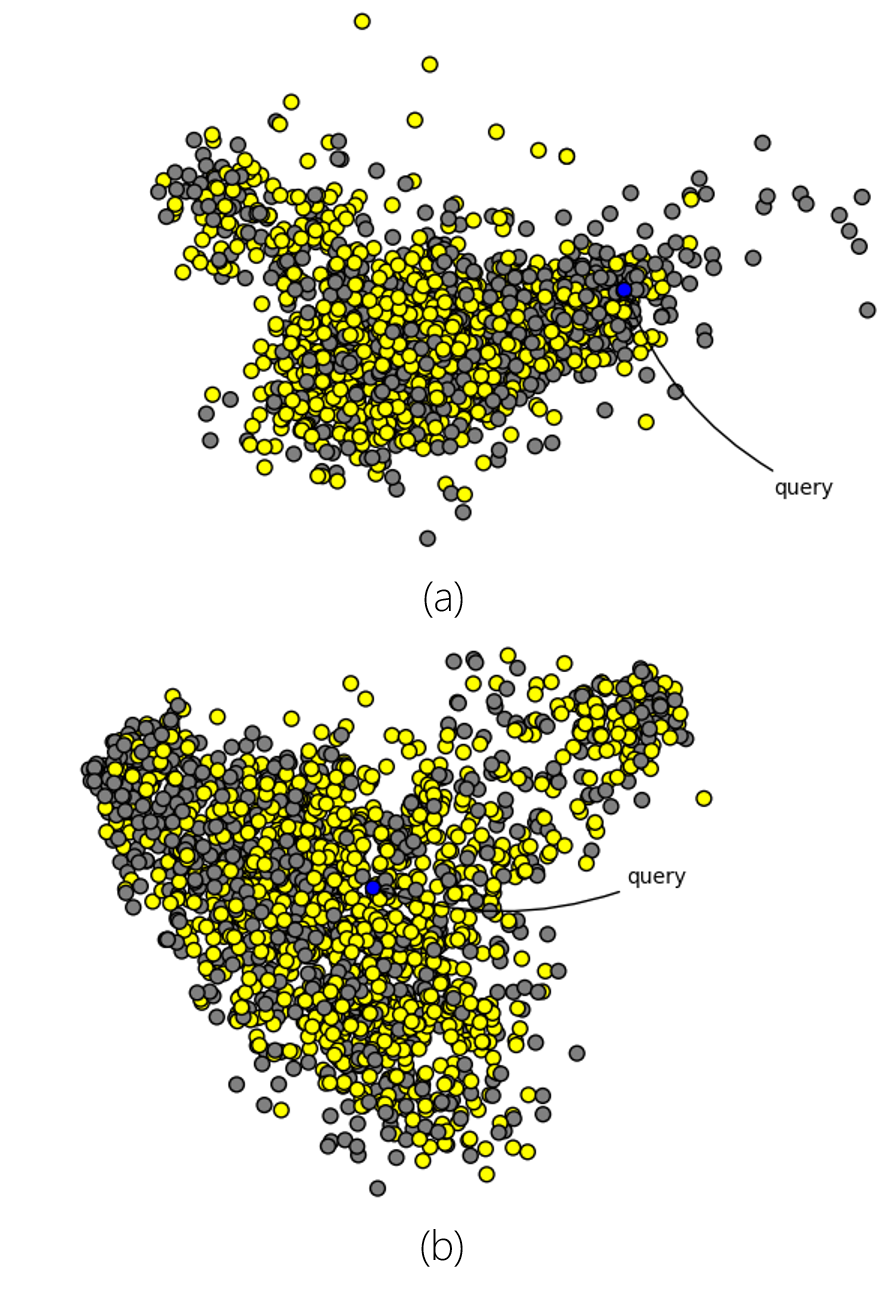}
\vspace{-1\baselineskip}
\caption{A two dimensional PCA projection of the 200-dimensional embeddings. Relevant documents are yellow, irrelevant documents are grey, and the query is blue. To visualize the results of multiple queries at once, before dimensionality reduction we centre query vectors at the origin and represent documents as the difference between the document vector and its query vector. (a) uses IN word vector centroids to represent both the query and the documents. (b) uses IN for the queries and OUT for the documents, and seems to have a higher density of relevant documents near the query.}
\vspace{-1\baselineskip}
\label{fig:PCA}
\end{figure}

\begin{table*}[t]
\renewcommand{\arraystretch}{1.1}\addtolength{\tabcolsep}{2.5pt}
\begin{center}
\caption{The nearest neighbours for the words "yale", "seahawks" and "eminem" according to the cosine similarity based on the IN-IN, OUT-OUT and IN-OUT vector comparisons for the different words in the vocabulary. These examples show that IN-IN and OUT-OUT cosine similarities are high for words that are similar by function or type (\emph{typical}), and the IN-OUT cosine similarities are high between words that often co-occur in the same query or document (\emph{topical}). The \emph{word2vec} model used here was trained on a query corpus with a vocabulary of 2,748,230 words.}
\label{tbl:results-nearestneighbors}
\vspace{.5\baselineskip}
\resizebox{.96\textwidth}{!}{
\begin{tabular}{c c c c c c c c c c c c c}
  \midrule
  & \textbf{yale} &  & & & & \textbf{seahawks} & & & & & \textbf{eminem} & \\
IN-IN & OUT-OUT & IN-OUT & & & IN-IN & OUT-OUT & IN-OUT & & & IN-IN & OUT-OUT & IN-OUT \\ \cline{1-3} \cline{6-8} \cline{11-13}
yale & yale & yale & & & seahawks & seahawks & seahawks & & & eminem & eminem & eminem \\
harvard & uconn& faculty & & & 49ers & broncos & highlights & & & rihanna & rihanna & rap \\
nyu & harvard & alumni & & & broncos & 49ers & jerseys & & & ludacris & dre & featuring \\
cornell & tulane & orientation & & & packers & nfl & tshirts & & & kanye & kanye & tracklist \\
tulane & nyu & haven & & & nfl & packers & seattle & & & beyonce & beyonce & diss \\
tufts & tufts & graduate & & & steelers & steelers & hats & & & 2pac & tupac & performs \\
  \bottomrule
\end{tabular}
}
\end{center}
\end{table*}

Identifying relevant documents for a given query is a core challenge for Web search. For large-scale search engines, it is possible to identify a very small set of pages that can answer a good proportion of queries \cite{baeza:www2015}. For such popular pages, clicks and hyperlinks may provide sufficient ranking evidence and it may not be important to match the query against the body text. However, in many Web search scenarios such query-content matching is crucial. If new content is available, the new and updated documents may not have click evidence or may have evidence that is out of date. For new or tail queries, there may be no memorized connections between the queries and the documents. Furthermore, many search engines and apps have a relatively smaller number of users, which limits their ability to answer queries based on memorized clicks. There may even be insufficient behaviour data to learn a click-based embedding \cite{huang2013learning} or a translation model \cite{gao2011clickthrough, Jones:etal:WWW2006}. In these cases it is crucial to model the relationship between the query and the document content, without click data.
 
When considering the relevance of document body text to a query, the traditional approach is to count repetitions of query terms in the document. Different transformation and weighting schemes for those counts lead to a variety of possible TF-IDF ranking features. One theoretical basis for such features is the probabilistic model of information retrieval, which has yielded the very successful TF-IDF formulation BM25\cite{robertson2009probabilistic}. As noted by \citet{robertson2004understanding}, the probabilistic approach can be restricted to consider only the original query terms or it can automatically identify additional terms that are correlated with relevance. However, the basic commonly-used form of BM25 considers query terms only, under the assumption that non-query terms are less useful for document ranking.

In the probabilistic approach, the 2-Poisson model forms the basis for counting term frequency \cite{bookstein:JASIS1974, harter:JASIS1975, robertson:sigir1994poisson}. The stated goal is to distinguish between a document that is about a term and a document that merely mentions that term. These two types of documents have term frequencies from two different Poisson distributions, such that documents about the term tend to have higher term frequency than those that merely mention it. This explanation for the relationship between term frequency and aboutness is the basis for the TF function in BM25 \cite{robertson:sigir1994poisson}.

The new approach in this paper uses word occurrences as evidence of aboutness, as in the probabilistic approach. However, instead of considering term repetition as evidence of aboutness it considers the relationship between the query terms and all the terms in the document. For example, given a query term ``yale'', in addition to considering the number of times Yale is mentioned in the document, we look at whether related terms occur in the document, such as ``faculty'' and ``alumni''. Similarly, in a document about the Seahawks sports team one may expect to see the terms ``highlights'' and ``jerseys''. The occurrence of these related terms in sufficient numbers is a way to distinguish between documents that merely mention Yale or Seahawks and the documents that are about the university or about the sports team. 

With this motivation, in Section \ref{sec:model} we describe how the input and the output embedding spaces learned by a \emph{word2vec} model may be jointly particularly attractive for modelling the \emph{aboutness} aspect of document ranking. Table \ref{tbl:results-nearestneighbors} gives some anecdotal evidence of why this is true. If we look in the neighbourhood of the IN vector of the word ``yale'' then the other IN vectors that are close correspond to words that are functionally similar or of the same type, e.g., ``harvard'' and ``nyu''. A similar pattern emerges if we look at the OUT vectors in the neighbourhood of the OUT vector of ``yale''. On the other hand, if we look at the OUT vectors that are closest to the IN vector of ``yale'' we find words like ``faculty'' and ``alumni''. We use this property of the IN-OUT embeddings to propose a novel \emph{Dual Embedding Space Model} (DESM) for document ranking. Figure \ref{fig:PCA} further illustrates how in this Dual Embedding Space model, using the IN embeddings for the query words and the OUT embeddings for the document words we get a much more useful similarity definition between the query and the relevant document centroids.

The main contributions of this paper are,
\begin{itemize}
\itemsep-2pt
\item A novel Dual Embedding Space Model, with one embedding for query words and a separate embedding for document words, learned jointly based on an unlabelled text corpus.
\item We propose a document ranking feature based on comparing all the query words with all the document words, which is equivalent to comparing each query word to a centroid of the document word embeddings.
\item We analyse the positive aspects of the new feature, preferring documents that contain many words related to the query words, but also note the potential of the feature to have false positive matches.
\item We empirically compare the new approach to a single embedding and the traditional word counting features. The new approach works well on its own in a \emph{telescoping} setting, re-ranking the top documents returned by a commercial Web search engine, and in combination with word counting for a more general document retrieval task.
\end{itemize}

\section{Distributional Semantics for IR}
\label{sec:model}

In this section we first introduce the \textit{Continuous Bag-of-Words} (CBOW) model made popular by the software \textit{Word2Vec} \cite{mikolov2013efficient, mikolov2013distributed}. Then, inspired by our findings that distinctly different topic-based relationships can be found by using both the input and the output embeddings jointly -- the latter of which is usually discarded after training -- we propose the \emph{Dual Embedding Space Model} (DESM) for document ranking.

\subsection{Continuous Bag-of-Words}
While many word embedding models have been proposed recently, the \emph{Continuous Bag-of-Words} (CBOW) and the \emph{Skip-Gram} (SG) architectures proposed by \citet{mikolov2013distributed} are arguably the most popular (perhaps due to the popularity of the software \emph{Word2Vec}\footnote{https://code.google.com/p/word2vec/}, which implements both).  Although here we will concentrate exclusively on the CBOW model, our proposed IR ranking methodology is just as applicable to vectors produced by SG, as both models produce qualitatively and quantitatively similar embeddings.  

The CBOW model learns a word's embedding via maximizing the log conditional probability of the word given the context words occurring within a fixed-sized window around that word.  That is, the words in the context window serve as input, and from them, the model attempts to predict the center (missing) word.  For a formal definition, let $\mathbf{c}_{k} \in \mathbb{R}^{d}$ be a $d$-dimensional, real-valued vector representing the $k$th context word $c_{k}$ appearing in a $K-1$-sized window around an instance of word $w_{i}$, which is represented by a vector $\mathbf{w}_{i} \in \mathbb{R}^{d}$.  The model `predicts' word $w_{i}$ by adapting its representation vector such that it has a large inner-product with the mean of the context word vectors.  Training CBOW requires minimization of the following objective 

\begin{equation}
\begin{split}
\label{cbow_def} 
\mathcal{L}_{CBOW} &= \sum_{i=1}^{|\mathcal{D}|} - \log p(w_{i} | C_{K}) \\
&=  \sum_{i=1}^{|\mathcal{D}|} - \log \frac{e^{\mathbf{\overline{C}}_{K}^{T} \mathbf{w}_{i}}}{\sum_{v=1}^{V_{w}} e^{\mathbf{\overline{C}}_{K}^{T}  \mathbf{w}_{v}}},  \end{split}
\end{equation} 

where

\begin{equation}
\mathbf{\overline{C}}_{K} = \frac{1}{K-1}\sum_{i-K\le k \le i+K, k\ne i} \mathbf{c}_{k}
\end{equation}

and $\mathcal{D}$ represents the training corpus.  Notice that the probability is normalized by summing over all the vocabulary, which is quite costly when training on web-scale data.  To make CBOW scalable, \citet{mikolov2013distributed} proposed the following slightly altered \textit{negative sampling} objective:   

\begin{equation}
- \log p(w_{i} | C_{K}) \approx -\log \sigma(\mathbf{\overline{C}}_{K}^{T}\mathbf{w}_{i}) - \sum_{n = 1}^{N} \log \sigma(-\mathbf{\overline{C}}_{K}^{T}\mathbf{\hat w}_{n})
\end{equation}

where $\sigma$ is the Sigmoid function and $N$ is the number of negative sample words drawn either from the uniform or empirical distribution over the vocabulary.  All our experiments were performed with the negative sampling objective.

\begin{figure}[t]
\center
\vspace{1\baselineskip}
\includegraphics[width=2.8in]{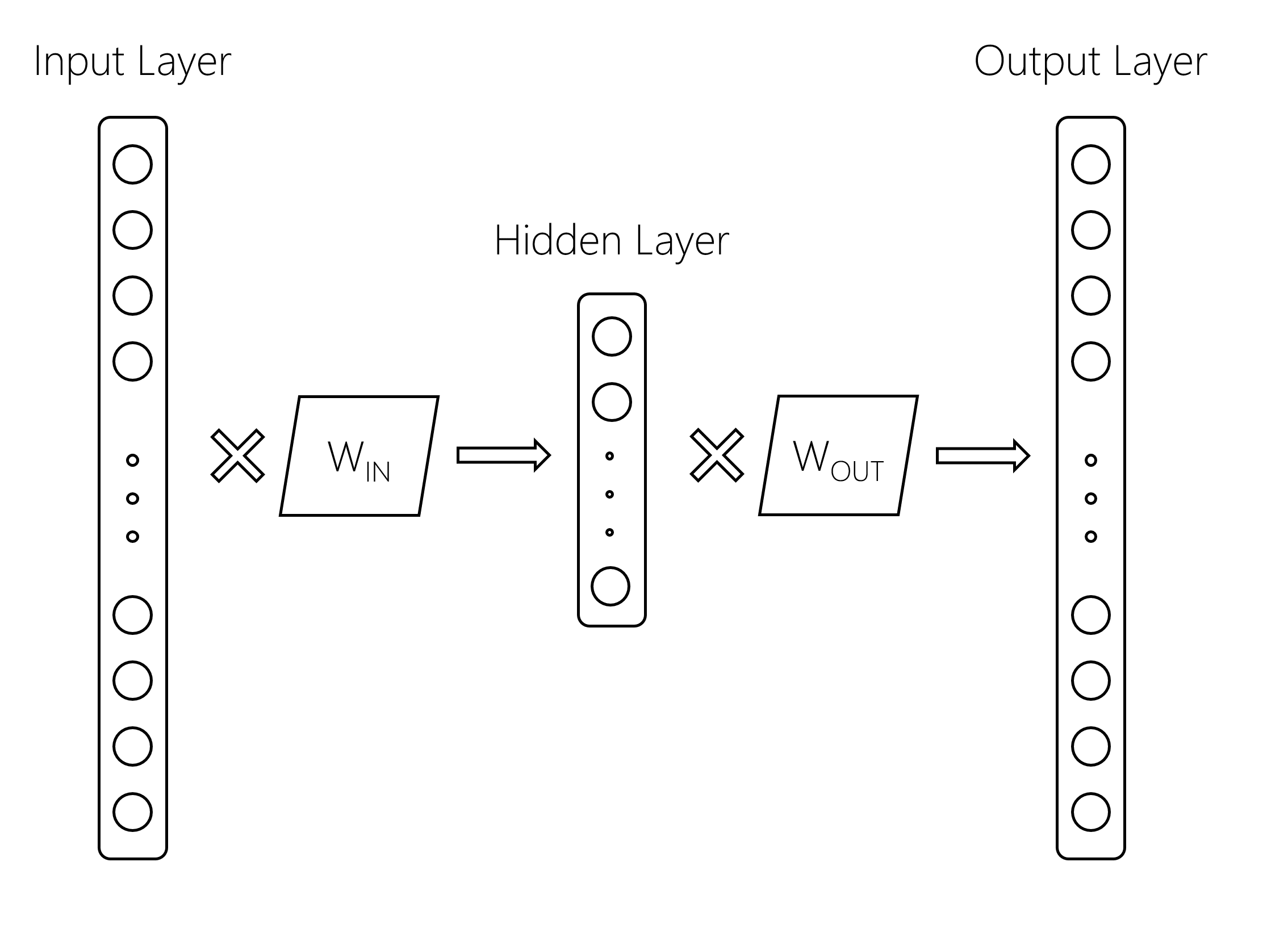}
\vspace{-.5\baselineskip}
\caption{The architecture of a word2vec (CBOW) model considering a single context word. $W_{IN}$ and $W_{OUT}$ are the two weight matrices learnt during training and corresponds to the IN and the OUT word embedding spaces of the model.}
\label{fig:architecture}
\end{figure}

A crucial detail often overlooked when using Word2Vec is that there are two different sets of vectors (represented above by $\mathbf{c}$ and $\mathbf{w}$ respectively and henceforth referred to as the \emph{IN} and \emph{OUT} embedding spaces), which correspond to the $\mathbf{W}_{IN}$ and $\mathbf{W}_{OUT}$ weight matrices in Figure \ref{fig:architecture}.  By default, Word2Vec discards $\mathbf{W}_{OUT}$ at the end of training and outputs only $\mathbf{W}_{IN}$.  Subsequent tasks determine word-to-word semantic relatedness by computing the cosine similarity:   

\begin{equation}
sim(c_{i}, c_{j}) = cos(\mathbf{c}_{i}, \mathbf{c}_{j}) = \frac{\mathbf{c}_{i}^{T}\mathbf{c}_{j}}{\norm{\mathbf{c}_{i}}\norm{\mathbf{c}_{j}}}
\end{equation}

\vspace{1\baselineskip}

\subsection{Dual Embedding Space Model}
\label{desm}

\begin{figure*}[t]
\center
\includegraphics[width=6.5in]{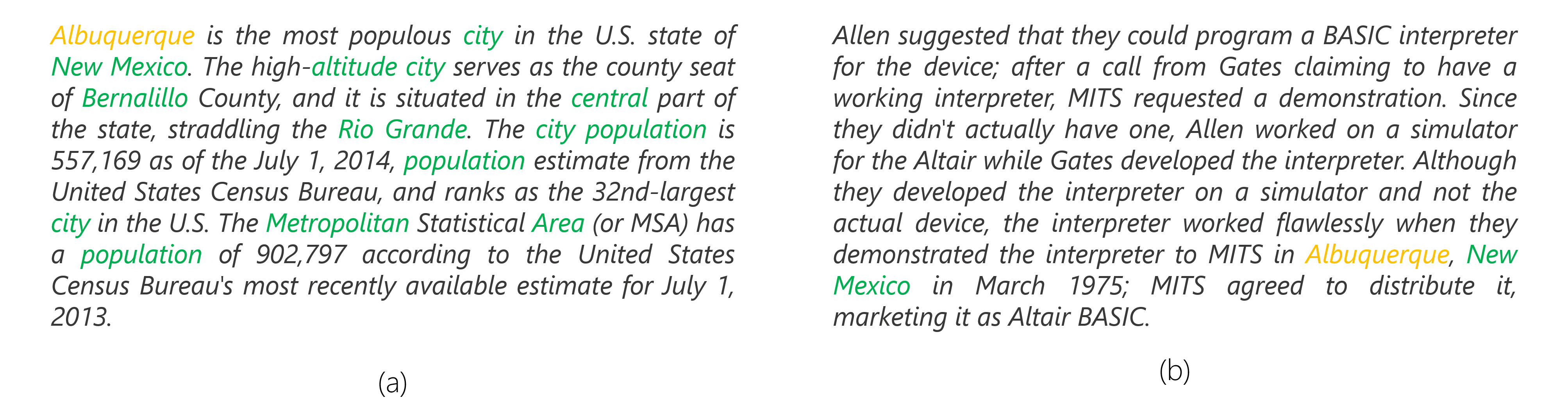}
\vspace{-1\baselineskip}
\caption{Two different passages from Wikipedia that mentions "Albuquerque" (highlighted in orange) exactly once. Highlighted in green are all the words that have an IN-OUT similarity score with the word "Albuquerque" above a fixed threshold (we choose -0.03 for this visualization) and can be considered as providing supporting evidence that (a) is \emph{about} Albuquerque, whereas (b) happens to only \emph{mention} the city.}
\vspace{-.8\baselineskip}
\label{fig:aboutness}
\end{figure*}

\begingroup
\begin{table*}[t]
\renewcommand{\arraystretch}{1.1}\addtolength{\tabcolsep}{.5pt}
\vspace{.5\baselineskip}
\begin{center}
\caption{A word perturbation analysis to show how the DESM collects evidence on the \emph{aboutness} of a document. The DESM models are more robust irrelevant terms. For example, when the word "giraffe" is replaced by the word "cambridge", the passage on giraffes is still scored low by the DESM for the query "cambridge" because it finds low supporting evidence from the other words in the passage. However, the DESM confuses the passage about Oxford to be relevant for the query "cambridge" because it detects a high number of similar words in the passage that frequently co-occur with the word "Cambridge".}
\vspace{1\baselineskip}
\label{tbl:perturb}
\resizebox{.99\textwidth}{!}{
\begin{tabular}{| p{2cm}| p{13cm}| c | c | c |}
 \hline
\multicolumn{5}{|c|}{Query: "cambridge"} \\
 \hline
 & & DESM & DESM & Term \\
Passage type & Passage text  & (IN-OUT) & (IN-IN) & Frequency \\
 &  & Score & Score & Count \\
 \hline
Passage about Cambridge & The city of Cambridge is a university city and the county town of Cambridgeshire, England. It lies in East Anglia, on the River Cam, about 50 miles (80 km) north of London. According to the United Kingdom Census 2011, its population was 123,867 (including 24,488 students). This makes Cambridge the second largest city in Cambridgeshire after Peterborough, and the 54th largest in the United Kingdom. There is archaeological evidence of settlement in the area during the Bronze Age and Roman times; under Viking rule Cambridge became an important trading centre. The first town charters were granted in the 12th century, although city status was not conferred until 1951. & -0.062 & 0.120 & 5 \\
 \hline
Passage about Oxford & Oxford is a city in the South East region of England and the county town of Oxfordshire. With a population of 159,994 it is the 52nd largest city in the United Kingdom, and one of the fastest growing and most ethnically diverse. Oxford has a broad economic base. Its industries include motor manufacturing, education, publishing and a large number of information technology and science-based businesses, some being academic offshoots. The city is known worldwide as the home of the University of Oxford, the oldest university in the English-speaking world. Buildings in Oxford demonstrate examples of every English architectural period since the arrival of the Saxons, including the mid-18th-century Radcliffe Camera. Oxford is known as the city of dreaming spires, a term coined by poet Matthew Arnold. & -0.070 & 0.107 & 0 \\
 \hline
Passage about giraffes & The giraffe (Giraffa camelopardalis) is an African even-toed ungulate mammal, the tallest living terrestrial animal and the largest ruminant. Its species name refers to its camel-like shape and its leopard-like colouring. Its chief distinguishing characteristics are its extremely long neck and legs, its horn-like ossicones, and its distinctive coat patterns. It is classified under the family Giraffidae, along with its closest extant relative, the okapi. The nine subspecies are distinguished by their coat patterns. The giraffe's scattered range extends from Chad in the north to South Africa in the south, and from Niger in the west to Somalia in the east. Giraffes usually inhabit savannas, grasslands, and open woodlands. & -0.102 & 0.011 & 0 \\
 \hline
Passage about giraffes, but the word "giraffe" is replaced by the word "Cambridge" & The cambridge (Giraffa camelopardalis) is an African even-toed ungulate mammal, the tallest living terrestrial animal and the largest ruminant. Its species name refers to its camel-like shape and its leopard-like colouring. Its chief distinguishing characteristics are its extremely long neck and legs, its horn-like ossicones, and its distinctive coat patterns. It is classified under the family Giraffidae, along with its closest extant relative, the okapi. The nine subspecies are distinguished by their coat patterns. The cambridge's scattered range extends from Chad in the north to South Africa in the south, and from Niger in the west to Somalia in the east. giraffes usually inhabit savannas, grasslands, and open woodlands. & -0.094 & 0.033 & 3 \\
 \hline
Passage about Cambridge, but the word "Cambridge" is replaced by the word "giraffe" & The city of Giraffe is a university city and the county town of Cambridgeshire, England. It lies in East Anglia, on the River Cam, about 50 miles (80 km) north of London. According to the United Kingdom Census 2011, its population was 123,867 (including 24,488 students). This makes Giraffe the second largest city in Cambridgeshire after Peterborough, and the 54th largest in the United Kingdom. There is archaeological evidence of settlement in the area during the Bronze Age and Roman times; under Viking rule Giraffe became an important trading centre. The first town charters were granted in the 12th century, although city status was not conferred until 1951. & -0.076 & 0.088 & 0 \\
 \hline
\end{tabular}
}
\end{center}
\vspace{.5\baselineskip}
\end{table*}
\endgroup

A key challenge for term-matching based retrieval is to distinguish whether a document merely references a term or is about that entity.  See Figure \ref{fig:aboutness} for a concrete example of two passages that contain the term "Albuquerque" an equal number of times although only one of the passages  is \emph{about} that entity. The presence of the words like "population" and "metropolitan" indicate that the left passage is about Albuquerque, whereas the passage on the right just \emph{mentions} it.  However, these passages would be indistinguishable under term counting.  The semantic similarity of non-matched terms (i.e. the words a TF feature would overlook) are crucial for inferring a document's topic of focus--its \emph{aboutness}.   

Due to its ability to capture word co-occurrence (i.e. perform missing word prediction), CBOW is a natural fit for modelling the \emph{aboutness} of a document.  The learnt embedding spaces contain useful knowledge about the \emph{distributional} properties of words, allowing, in the case of Figure \ref{fig:aboutness}, an IR system to recognize the city-related terms in the left document.  With this motivation, we define a simple yet, as we will demonstrate, effective ranking function we call the \emph{Dual Embedding Space Model}:

\vspace{-1\baselineskip}
\begin{equation}
\begin{split}
\label{desm_def}
DESM(Q, D) &= \frac{1}{|Q|} \sum_{q_{i} \in Q}  \frac{\mathbf{q}_{i}^{T} \mathbf{\overline{D}}}{\norm{\mathbf{q}_{i}}\norm{\mathbf{\overline{D}}}},
\end{split}
\end{equation}
\vspace{-1\baselineskip}

where

\vspace{-1\baselineskip}
\begin{equation}
\begin{split}
\label{desm_doc_centroid}
\mathbf{\overline{D}} &= \frac{1}{|D|} \sum_{\mathbf{d}_{j} \in D}  \frac{ \mathbf{d}_{j}} {\norm{\mathbf{d}_{j}}} \end{split}
\end{equation}
\vspace{-1\baselineskip}

Here $\mathbf{\overline{D}}$ is the centroid of all the normalized vectors for the words in the document serving as a single embedding for the whole document. In this formulation of the DESM, the document embeddings can be pre-computed, and at the time of ranking, we only need to sum the score contributions across the query terms. We expect that the ability to pre-compute a single document embedding is a very useful property when considering runtime efficiency.

\vspace{.5\baselineskip}

\paragraph*{IN-IN vs. IN-OUT}
\citet{hill2014not} noted, "Not all neural embeddings are born equal". As previously mentioned, the CBOW (and SG) model contains two separate embedding spaces (IN and OUT) whose interactions capture additional distributional semantics of words that are not observable by considering any of the two embeddings spaces in isolation. Table \ref{tbl:results-nearestneighbors} illustrates clearly how the CBOW model "pushes" the IN vectors of words closer to the OUT vectors of other words that they commonly co-occur with. In doing so, words that appear in similar contexts get pushed closer to each other within the IN embedding space (and also within the OUT embedding space). Therefore the IN-IN (or the OUT-OUT) cosine similarities are higher for words that are \emph{typically} (by type or by function) similar, whereas the IN-OUT cosine similarities are higher for words that co-occur often in the training corpus (\emph{topically} similar). This gives us at least two variants of the DESM, corresponding to retrieval in the IN-OUT space or the IN-IN space\footnote{It is also possible to define $DESM_{OUT-OUT}$ and $DESM_{OUT-IN}$, but based on limited experimentation we expect them to behave similar to $DESM_{IN-IN}$ and $DESM_{IN-OUT}$, respectively.}.

\vspace{-1\baselineskip}
\begin{equation}
\label{desm_def_inout}
\begin{split}
DESM_{IN-OUT}(Q, D) = \frac{1}{|Q|} \sum_{q_{i} \in Q}  \frac{q_{IN, i}^{T} \overline{D_{OUT}}}{\norm{q_{IN, i}}\norm{\overline{D_{OUT}}}}
\end{split}
\end{equation}
\vspace{-1\baselineskip}

\begin{equation}
\label{desm_def_inin}
\begin{split}
DESM_{IN-IN}(Q, D) = \frac{1}{|Q|} \sum_{q_{i} \in Q}  \frac{q_{IN, i}^{T} \overline{D_{IN}}}{\norm{q_{IN, i}}\norm{\overline{D_{IN}}}}
\end{split}
\end{equation}
\vspace{-1\baselineskip}

In Section \ref{sec:results}, we show that the $DESM_{IN-OUT}$ is a better indication of \emph{aboutness} than BM25, because of its knowledge of the word distributional properties, and $DESM_{IN-IN}$, since \emph{topical} similarity is a better indicator of \emph{aboutness} than \emph{typical} similarity.

\paragraph*{Modelling document aboutness}
\label{sec:aboutness}
We perform a simple word perturbation analysis to illustrate how the DESM can collect evidence on document \emph{aboutness} from both matched and non-matched terms in the document. In Table \ref{tbl:perturb}, we consider five small passages of text. The first three passages are about Cambridge, Oxford and giraffes respectively. The next two passages are generated by replacing the word "giraffe" by the word "Cambridge" in the passage about giraffes, and vice versa.

We compute the $DESM_{IN-OUT}$ and the $DESM_{IN-IN}$ scores along with the term frequencies for each of these passages for the query term "cambridge". As expected, all three models score the passage about Cambridge highly. However, unlike the term frequency feature, the DESM seem robust towards \emph{keyword stuffing}\footnote{\url{https://en.wikipedia.org/wiki/Keyword_stuffing}}, at least in this specific example where we replace the word "giraffe" with "cambridge" in the passage about giraffes, but the DESMs still score the passage relatively low. This is exactly the kind of evidence that we expect the DESM to capture that may not be possible by simple term counting.

On the other hand, both the DESMs score the passage about Oxford very highly. This is expected because both these passages contain many words that are likely to co-occur with the word "cambridge" in the training corpus. This implies that the DESM features are very susceptible to false positive matches and can only be used either in conjunction with other document ranking features, such as TF-IDF, or for re-ranking a smaller set of candidate documents already deemed at least somewhat relevant. This is similar to the \emph{telescoping} evaluation setup described by \citet{matveeva:sigir06:telescoping}, where multiple nested rankers are used to achieve better retrieval performance over a single ranker. At each stage of telescoping, a ranker is used to reduce the set of candidate documents that is passed on to the next. Improved performance is possible because the ranker that sees only top-scoring documents can specialize in handling such documents, for example by using different feature weights. In our experiments, we will see the DESM to be a poor standalone ranking signal on a larger set of documents, but performs significantly better against the BM25 and the LSA baselines once we reach a small high-quality candidate document set. This evaluation strategy of focusing at ranking for top positions is in fact quite common and has been used by many recent studies (e.g., \cite{gao2011clickthrough, huang2013learning}).

\paragraph*{Dot product vs. cosine similarity}
In the DESM formulation (Equation \ref{desm_def}) we compute the cosine similaritiy between every query word and the normalized document centroid. The use of cosine similarity (as opposed to, say, dot-product) is motivated by several factors. Firstly, much of the existing literature\cite{mikolov2013efficient, mikolov2013distributed} on CBOW and SG uses cosine similarity and normalized unit vectors (for performing vector algebra for word analogies). As the cosine similarity has been shown to perform well in practice in these embedding spaces we adopt the same strategy here.

A secondary justification can be drawn based on the observations made by \citet{Wilson2015Embed} that the length of the non-normalized word vectors has a direct relation to the frequency of the word. In information retrieval (IR), it is well known that frequently occurring words are ineffective features for distinguishing relevant documents from irrelevant ones. The inverse-document frequency weighting is often used in IR to capture this effect. By normalizing the word vectors in the document before computing the document centroids, we are counteracting the extra influence frequent words would have on the sum.

\paragraph*{Training corpus}
Our CBOW model is trained on a query corpus\footnote{We provide the IN and OUT word embeddings trained using word2vec on the Bing query corpus at \url{http://research.microsoft.com/projects/DESM}.} consisting of 618,644,170 queries and a vocabulary size of 2,748,230 words.  The queries are sampled from Bing's large scale search logs from the period of August 19, 2014 to August 25, 2014. We repeat all our experiments using another CBOW model trained on a corpus of document body text with 341,787,174 distinct sentences sampled from the Bing search index and a corresponding vocabulary size of 5,108,278 words. Empirical results on the performance of both the models are presented in Section \ref{sec:results}.

\paragraph*{Out-of-vocabulary (OOV) words}
One of the challenges of the embedding models is that they can only be applied to a fixed size vocabulary. It is possible to explore different strategies to deal with out-of-vocab (OOV) words in the Equation \ref{desm_def} \footnote{In machine translation there are examples of interesting strategies to handle out-of-vocabulary words (e.g., \cite{luong2015addressing})}. But we leave this for future investigation and instead, in this paper, all the OOV words are ignored for computing the DESM score, but not for computing the TF-IDF feature, a potential advantage for the latter.

\paragraph*{Document length normalization}
In Equation \ref{desm_def} we normalize the scores linearly by both the query and the document lengths. While more sophisticated length normalization strategies, such as pivoted document length normalization \cite{singhal1996pivoted}, are reasonable, we leave this also for future work.

\subsection{The Mixture Model}
\label{sec:mixturemodel}

The DESM is a weak ranker and while it models some important aspects of document ranking, our experiments will show that it's effective only at ranking at high positions (i.e. documents we already know are at least somewhat relevant). We are inspired by previous work in neural language models, for example by \citet{bengio2003neural}, which demonstrates that combining a neural model for predicting the next word with a more traditional counting-based language model is effective because the two models make different kinds of mistakes. Adopting a similar strategy we propose a simple and intuitive mixture model combining DESM with a term based feature, such as BM25, for the non-telescoping evaluation setup described in Section \ref{sec:expsetup}.

We define the mixture model MM(Q, D) as,
\vspace{.5\baselineskip}
\begin{equation}
\begin{split}
\label{mixturemodel} 
MM(Q, D) = \alpha DESM(Q, D) + (1 - \alpha )BM25(Q, D) \\
\alpha \in \mathbb{R}, 0 \leq \alpha \leq 1
\end{split}
\end{equation}

To choose the appropriate value for $\alpha$, we perform a parameter sweep between zero and one at intervals of 0.01 on the implicit feedback based training set described in Section \ref{sec:datasets}.

\section{Experiments}
\label{sec:experiments}

\begingroup
\begin{table*}[t]
\renewcommand{\arraystretch}{1.1}\addtolength{\tabcolsep}{2.5pt}
\begin{center}
\caption{NDCG results comparing the $DESM_{IN-OUT}$ with the BM25 and the LSA baselines. The $DESM_{IN-OUT}$ performs significantly better than both the BM25 and the LSA baselines at all rank positions. It also performs better than the $DESM_{IN-IN}$ on both the evaluation sets. The DESMs using embeddings trained on the query corpus also performs better than if trained on document body text. The highest NDCG values for every column is highlighted in bold and all the statistically significant (p < 0.05) differences over the BM25 baseline are marked with the asterisk (*).}
\label{tbl:results-main}
\vspace{.5\baselineskip}
\resizebox{.96\textwidth}{!}{
\begin{tabular}{ l l l l l l l l l}
  \toprule
 & & \multicolumn{3}{c}{Explicitly Judged Test Set} & & \multicolumn{3}{c}{Implicit Feedback based Test Set} \\ \cline{3-5} \cline{7-9}
 & &  NDCG$\mathcal{@}$1 & NDCG$\mathcal{@}$3 &  NDCG$\mathcal{@}$10 & & NDCG$\mathcal{@}$1 & NDCG$\mathcal{@}$3 &  NDCG$\mathcal{@}$10\\
  \midrule
BM25 & & 23.69 & 29.14 & 44.77 & & 13.65 & 27.41 & 49.26  \\
LSA & & 22.41* & 28.25* & 44.24* & & 16.35* & 31.75* & 52.05* \\
DESM (IN-IN, trained on body text) & & 23.59 & 29.59 & 45.51* & & 18.62* & 33.80* & 53.32* \\
DESM (IN-IN, trained on queries) & & 23.75 & 29.72 & 46.36* & & 18.37* & 35.18* & 54.20* \\
DESM (IN-OUT, trained on body text) & & 24.06 & 30.32* & 46.57* & & 19.67* & 35.53* & 54.13* \\
DESM (IN-OUT, trained on queries) & & \textbf{25.02*} & \textbf{31.14*} & \textbf{47.89*} & & \textbf{20.66*} & \textbf{37.34*} & \textbf{55.84*} \\
  \bottomrule
\end{tabular}
}
\end{center}
\vspace{-1\baselineskip}
\end{table*}
\endgroup

We compare the retrieval performance of DESM against BM25, a traditional count-based method, and Latent Semantic Analysis (LSA), a traditional vector-based method. We conduct our evaluations on two different test sets (explicit and implicit relevance judgements) and under two different experimental conditions (a large collection of documents and a telescoped subset). 

\subsection{Datasets}
\label{sec:datasets}
All the datasets that are used for this study are sampled from Bing's large scale query logs. The body text for all the candidate documents are extracted from Bing's document index.

\paragraph*{Explicitly judged test set}
This evaluation set consists of 7,741 queries randomly sampled from Bing's query logs from the period of October, 2014 to December, 2014. For each sampled query, a set of candidate documents is constructed by retrieving the top results from Bing over multiple scrapes during a period of a few months. In total the final evaluation set contains 171,302 unique documents across all queries which are then judged by human evaluators on a five point relevance scale (Perfect, Excellent, Good, Fair and Bad).

\paragraph*{Implicit feedback based test set}
This dataset is sampled from the Bing logs from the period of the September 22, 2014 to September 28, 2014. The dataset consists of the search queries submitted by the user and the corresponding documents that were returned by the search engine in response. The documents are associated with a binary relevance judgment based on whether the document was clicked by the user. This test set contains 7,477 queries and the 42,573 distinct documents.

\paragraph*{Implicit feedback based training set}
This dataset is sampled exactly the same way as the previous test but from the period of September 15, 2014 to September 21, 2014 and has 7,429 queries and 42,253 distinct documents. This set is used for tuning the parameters for the BM25 baseline and the mixture model.

\subsection{Experiment Setup}
\label{sec:expsetup}

We perform two distinct sets of evaluations for all the experimental and baseline models. In the first experiment, we consider all documents retrieved by Bing (from the online scrapes in the case of the explicitly judged set or as recorded in the search logs in the case of the implicit feedback based sets) as the candidate set of documents to be re-ranked for each query.  The fact that each of the documents were retrieved by the search engine implies that they are all at least marginally relevant to the query.  Therefore, this experimental design isolates performance at the top ranks. As mentioned in Section \ref{sec:aboutness}, there is a parallel between this experiment setup and the \emph{telescoping} \cite{matveeva:sigir06:telescoping} evaluation strategy, and has been used often in recent literature (e.g., \cite{huang2013learning, shen2014learning}).  Note that by having a strong retrieval model, in the form of the Bing search engine, for first stage retrieval enables us to have a high confidence candidate set and in turn ensures reliable comparison with the baseline BM25 feature.

In our \emph{non-telescoped} experiment, we consider every distinct document in the test set as a candidate for every query in the same dataset. This setup is more in line with the traditional IR evaluation methodologies, where the model needs to retrieve the most relevant documents from a single large document collection. Our empirical results in Section \ref{sec:results} will show that the DESM model is a strong re-ranking signal, but as a standalone ranker, it is prone to false positives.  Yet, when we mix our neural model (DESM) with a counting based model (BM25), good performance is achieved.

For all the experiments we report the normalized discounted cumulative gain (NDCG) at different rank positions as a measure of performance for the different models under study.

\subsection{Baseline models}
\label{sec:baselines}

We compare the DESM models to a term-matching based baseline, in \emph{BM25}, and a vector space model baseline, in \emph{Latent Semantic Analysis} (LSA)\cite{deerwester1990indexing}. For the BM25 baseline we use the values of 1.7 for the $k_1$ parameter and 0.95 for the $b$ parameter based on a parameter sweep on the implicit feedback based training set. The LSA model is trained on the body text of 366,470 randomly sampled documents from Bing's index with a vocabulary size of 480,608 words. Note that unlike the word2vec models that train on word co-occurrence data, the LSA model by default trains on a word-document matrix.

\section{Results}
\label{sec:results}

\begingroup
\begin{table*}[t]
\renewcommand{\arraystretch}{1.1}\addtolength{\tabcolsep}{2.5pt}
\begin{center}
\caption{Results of NDCG evaluations under the non-telescoping settings. Both the DESM and the LSA models perform poorly in the presence of random irrelevant documents in the candidate set. The mixture of $DESM_{IN-OUT}$ with BM25 achieves the best NDCG. The best NDCG values are highlighted per column in bold and all the statistically significant (p < 0.05) differences with the BM25 baseline are indicated by the asterisk (*)}.
\label{tbl:results-main-distractors}
\resizebox{.96\textwidth}{!}{
\begin{tabular}{ l l l l l l l l l}
  \toprule
 & & \multicolumn{3}{c}{Explicitly Judged Test Set} & & \multicolumn{3}{c}{Implicit Feedback based Test Set} \\ \cline{3-5} \cline{7-9}
 & &  NDCG$\mathcal{@}$1 & NDCG$\mathcal{@}$3 &  NDCG$\mathcal{@}$10 & & NDCG$\mathcal{@}$1 & NDCG$\mathcal{@}$3 &  NDCG$\mathcal{@}$10\\
  \midrule
BM25 & & 21.44 & 26.09 & 37.53 & & 11.68 & 22.14 & 33.19 \\
LSA & & 04.61* & 04.63* & 04.83* & & 01.97* & 03.24* & 04.54* \\
DESM (IN-IN, trained on body text) & & 06.69* & 06.80* & 07.39* & & 03.39* & 05.09* & 07.13* \\
DESM (IN-IN, trained on queries) & & 05.56* & 05.59* & 06.03* & & 02.62* & 04.06* & 05.92* \\
DESM (IN-OUT, trained on body text) & & 01.01* & 01.16* & 01.58* & & 00.78* & 01.12* & 02.07* \\
DESM (IN-OUT, trained on queries) & & 00.62* & 00.58* & 00.81* & & 00.29* & 00.39* & 01.36* \\
BM25 + DESM (IN-IN, trained on body text) & & 21.53 & 26.16 & 37.48 & & 11.96 & 22.58* & 33.70* \\
BM25 + DESM (IN-IN, trained on queries) & & \textbf{21.58} & 26.20 & 37.62 & & 11.91 & 22.47* & 33.72* \\
BM25 + DESM (IN-OUT, trained on body text) & & 21.47 & 26.18 & 37.55 & & 11.83 & 22.42* & 33.60* \\
BM25 + DESM (IN-OUT, trained on queries) & & 21.54 & \textbf{26.42*} & \textbf{37.86*} & & \textbf{12.22*} & \textbf{22.96*} & \textbf{34.11*} \\
  \bottomrule
\end{tabular}
}
\end{center}
\vspace{-1\baselineskip}
\end{table*}
\endgroup

\begin{figure*}[t]
\center
\includegraphics[width=6.5in]{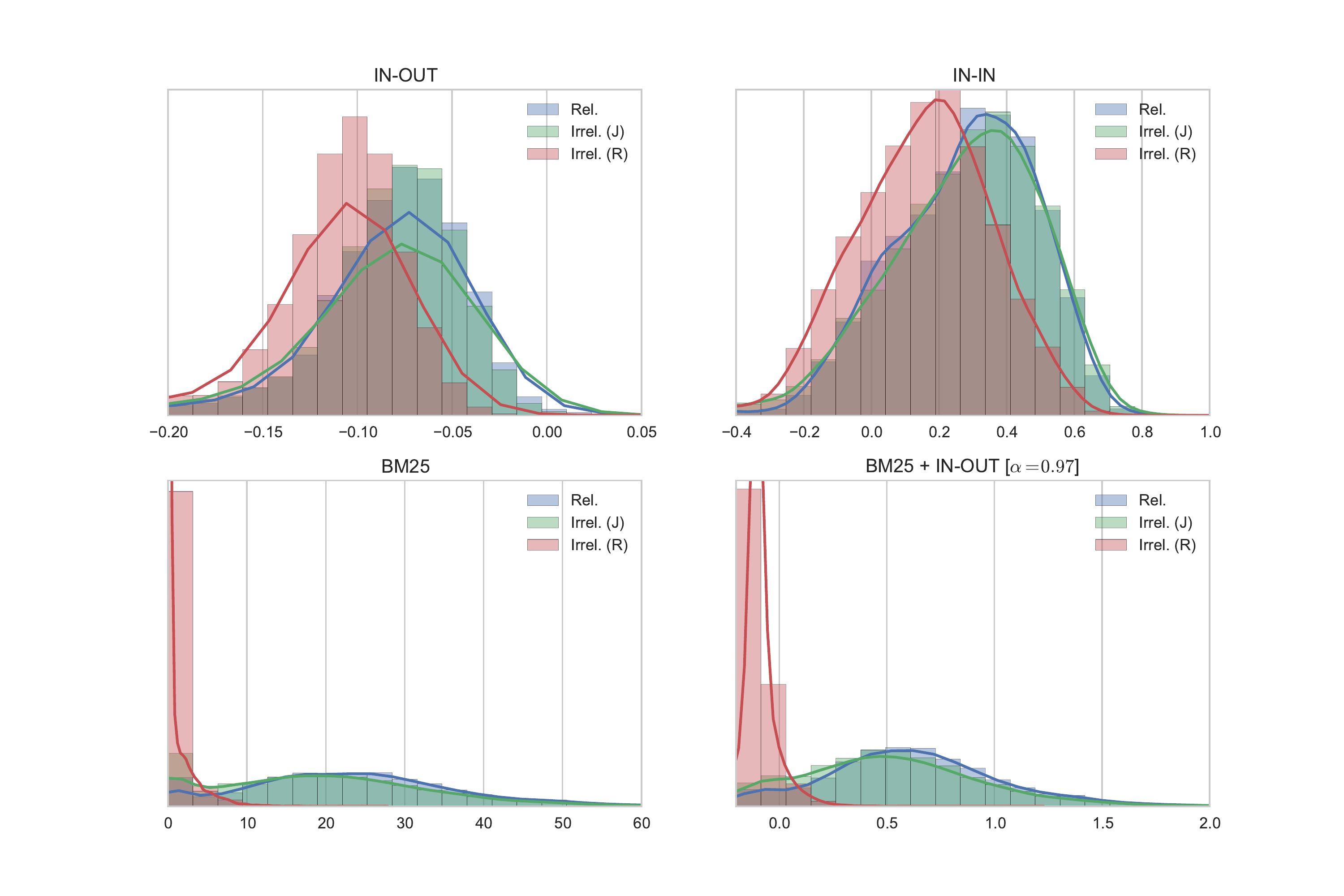}
\vspace{-.5\baselineskip}
\caption{Feature distributions over three sets of documents: \textbf{Rel.} retrieved by Bing and judged relevant, \textbf{Irrel. (J)} retrieved by Bing and judged irrelevant, and \textbf{Irrel. (R)} random documents not retrieved for this query. Our telescoping evaluation setup only uses the first two sets, whose distributions are quite close in all four plots. IN-OUT may have the greatest difference between Rel. and Irrel. (J), which corresponds to its good telescoping NDCG results. BM25 is far superior at separating Irrel. (R) results from the rest, which explains the success of BM25 and mixture models in non-telescoping evaluation.}
\vspace{-1\baselineskip}
\label{fig:histograms}
\end{figure*}

\begin{figure*}[t]
\center
\includegraphics[width=6.5in]{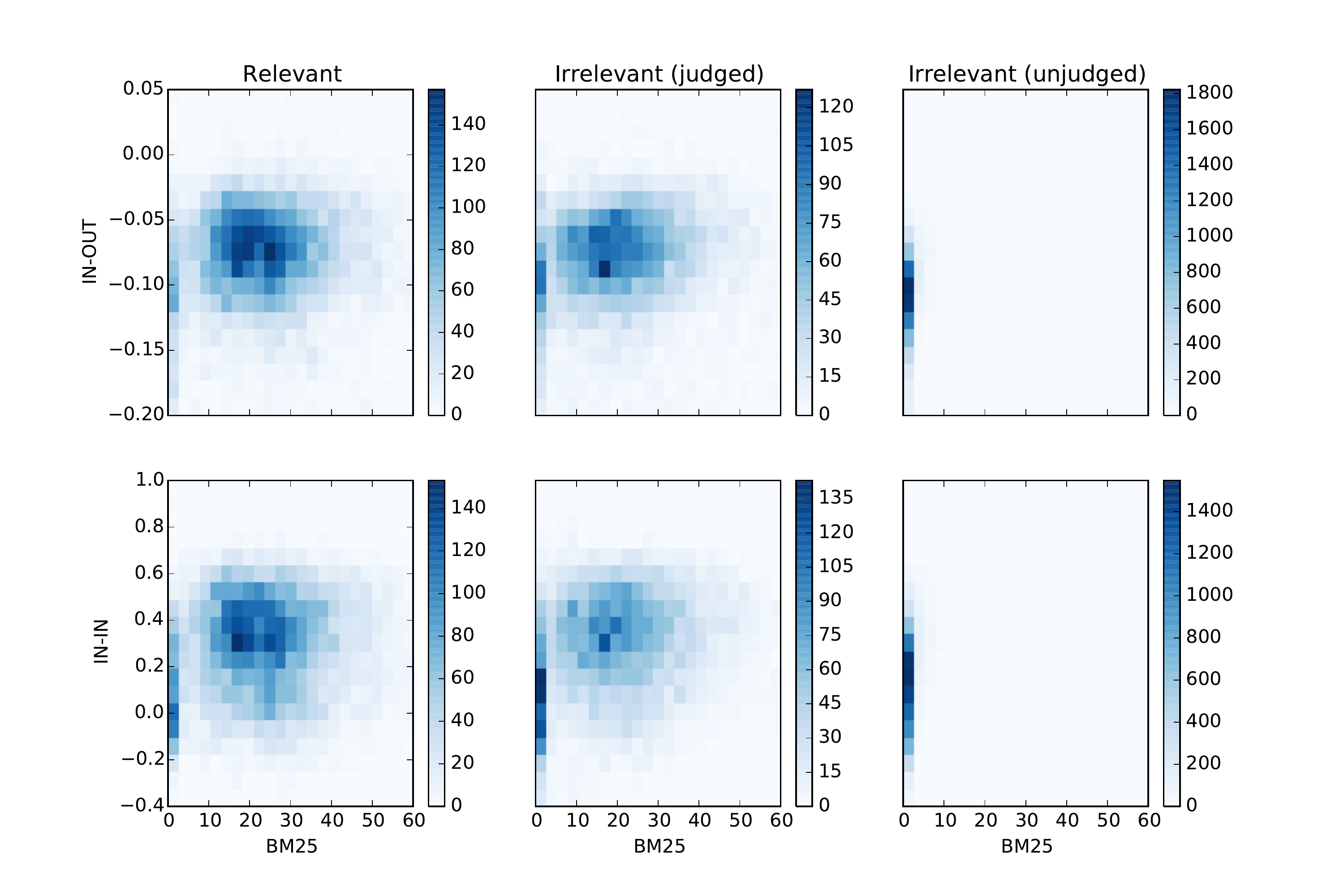}
\vspace{-1\baselineskip}
\caption{Bivariate analysis of our lexical matching and neural word embedding features. On unjudged (random) documents, BM25 is very successful at giving zero score, but both IN-IN and IN-OUT give a range of scores. This explains their poor performance in non-telescoping evaluation. For the judged relevant and judged irrelevant sets, we see a range of cases where both types of feature fail. For example BM25 has both false positives, where an irrelevant document mentions the query terms, and false negatives, where a relevant document does not mention the query terms.}
\vspace{-1\baselineskip}
\label{fig:heatmaps}
\end{figure*}

Table \ref{tbl:results-main} shows the NCDG based performance evaluations under the \emph{telescoping} setup. On both the explicitly judged and the implicit feedback based test sets the $DESM_{IN-OUT}$ performs significantly better than the BM25 and the LSA baselines, as well as the $DESM_{IN-IN}$ model. Under the \emph{all documents as candidates} setup in Table \ref{tbl:results-main-distractors}, however, the DESMs (both IN-IN and IN-OUT) are clearly seen to not perform well as standalone document rankers. The mixture of $DESM_{IN-OUT}$ (trained on queries) and BM25 rectifies this problem and gives the best NDCG result under the non-telescoping settings and demonstrates a statistically significant improvement over the BM25 baseline.

Figure \ref{fig:histograms} illustrates that the $DESM_{IN-OUT}$ is the most discriminating feature for the relevant and the irrelevant documents retrieved by a first stage retrieval system. However, BM25 is clearly superior in separating out the random irrelevant documents in the candidate set. The mixture model, unsurprisingly, has the good properties from both the $DESM_{IN-OUT}$  and the BM25 models. Figure \ref{fig:heatmaps} shows the joint distribution of the scores from the different models which further reinforces these points and shows that the DESM and the BM25 models make different errors.

We do not report the results of evaluating the mixture models under the telescoping setup because tuning the $\alpha$ parameter under those settings on the training set results in the best performance from the standalone DESM models. Overall, we conclude that the DESM is primarily suited for ranking at top positions or in conjunction with other document ranking features.

Interestingly, under the telescoping settings, the LSA baseline also shows some (albeit small) improvement over the BM25 baseline on the implicit feedback based test set but a loss on the explicitly judged test set.

With respect to the CBOW's training data, the DESM models with the embeddings trained on the query corpus performs significantly better than the models trained on document body text across different configurations. We have a plausible hypothesis on why this happens. Users tend to choose the most significant terms that they expect to match in the target document to formulate their search queries. Therefore in the query corpus, one may say that, the less important terms from the document corpus has been filtered out. Therefore when training on the query corpus the CBOW model is more likely to see important terms within the context window compared to when trained on a corpus of document body text, which may make it a better training dataset for the Word2vec model.

\section{Related Work}
\label{sec:related}

\paragraph*{Term based IR} 
For an overview of lexical matching approaches for information retrieval, such as the vector space, probabilistic and language modelling approach, see \cite{manning2008introduction}. In Salton's classic vector space model \cite{salton1975vector} queries and documents are represented as sparse vectors in a vector space of dimensionality |V|, where V is the word vocabulary. Elements in the vector are non-zero if that term occurs. Documents can be ranked in descending order of cosine similarity with the query, although a wide variety of weighting and similarity functions are possible \cite{zobel1998exploring}.  In contrast to the classical vector space model, LSA\cite{deerwester1990indexing}, PLSA\cite{hofmann1999probabilistic} and LDA\cite{blei2003latent, wei2006lda} learn dense vector representations of much lower dimensionality. It has been suggested that these models perform poorly as standalone retrieval models \cite{atreya2011latent} unless combined with other TF-IDF like features. In our approach the query and documents are also low dimensional dense vectors. We learn 200-dimensional neural word embeddings, and generate document vectors as the centroids of all the word vectors. \citet{yan2013learning} suggested that term correlation data is less sparse than term-document matrix and hence may be more effective for training embeddings.

The probabilistic model of information retrieval leads to the development of the BM25 ranking feature \cite{robertson2009probabilistic}. The increase in BM25 as term frequency increases is justified according to the 2-Poisson model \cite{harter:JASIS1975,robertson:sigir1994poisson}, which makes a distinction between documents about a term and documents that merely mention that term. Those two types of document have term frequencies from two different Poisson distributions, which justifies the use of term frequency as evidence of aboutness. By contrast, the model introduced in this paper uses the occurrence of other related terms as evidence of aboutness. For example, under the 2-Poisson model a document about Eminem will tend to mention the term `eminem' repeatedly. Under our all-pairs vector model, a document about Eminem will tend to contain more related terms such as `rap', `tracklist' and `performs'. Our experiments show both notions of aboutness to be useful.

\paragraph*{Neural embeddings for IR}

The word embeddings produced by the CBOW and SG models have been shown to be surprisingly effective at capturing detailed semantics useful for various Natural Language Processing (NLP) and reasoning tasks, including word analogies \cite{mikolov2013efficient, mikolov2013distributed}. Recent papers have explored in detail the SG and CBOW training methodology \cite{goldberg2014word2vec, rong2014word2vec} and its connection to other approaches for learning word embeddings such as explicit vector space representations \cite{levy2014linguistic, levy2015improving}, matrix factorization \cite{pennington2014glove, levy2014neural, shi2014linking} and density-based representations \cite{vilnis2014word}.

\citet{baroni2014don} evaluated neural word embeddings against traditional word counting approaches and demonstrated the success of the former on a variety of NLP tasks. However, more recent works \cite{schnabel2015evaluation, hill2014not} have shown that there does not seem to be one embedding approach that is best for all tasks. This observation is similar to ours, where we note that IN-IN and IN-OUT model different kinds of word relationships. Although IN-IN, for example, works well for word analogy tasks \cite{mikolov2013efficient, mikolov2013distributed}, it might perform less effectively for other tasks, such as those in information retrieval. If so, instead of claiming that any one embedding captures ``semantics'', it is probably better to characterize embeddings according to which tasks they perform well on.

Our paper is not the first to apply neural word embeddings in IR. \citet{ganguly2015word} recently proposed a generalized language model for IR that incorporates IN-IN similarities. The similarities are used to expand and reweight the terms in each document, which seems to be motivated by intuitions similar to ours, where a term is reinforced if a similar terms occurs in the query. In their case, after greatly expanding the document vocabulary, they perform retrieval based on word occurrences rather than in an embedding space.  Word embeddings have also been studied in other IR contexts such as term reweighting \cite{zheng2015learning}, cross-lingual retrieval \cite{vulic2015monolingual, zou2013bilingual, gupta2014query} and short-text similarity \cite{kenter15short}. Beyond word co-occurrence, recent studies have also explored learning text embeddings from clickthrough data \cite{huang2013learning, shen2014learning}, session data \cite{grbovic2015context, grbovic2015search, mitraexploring}, query prefix-suffix pairs \cite{mitra2015suffixrank}, via auto-encoders \cite{salakhutdinov2009semantic}, and for sentiment classification \cite{tang2014learning} and for long text\cite{le2014distributed}.

\section{Discussion and Conclusion}
\label{sec:conclusion}

This paper motivated and evaluated the use neural word embeddings to gauge a document's \emph{aboutness} with respect to a query.  Mapping words to points in a shared semantic space allows a query term to be compared against all terms in the document, providing for a refined relevance scoring.  We formulate a \textit{Dual Embedding Space Model} (DESM) that leverages the often discarded output embeddings learned by the CBOW model.   Our model exploits a novel use of both the input and output embeddings to capture topic-based semantic relationships.  The examples in Table\ref{tbl:results-nearestneighbors} show that drastically different nearest neighbors can be found by using proximity in the IN-OUT vs the IN-IN spaces.  We have demonstrated through intuition and large-scale experimentation that ranking via proximity in IN-OUT space is better for retrieval than IN-IN based rankers.  This finding emphasizes that usage of the CBOW and SG models is application dependent and that quantifying semantic relatedness via cosine similarity in IN space should not be a default practice.  

We have also identified and investigated a failure of embedding-based ranking: performance is highly dependent on the relevancy of the initial candidate set of documents to be ranked.  While stand-alone DESM clearly bests BM25 and LSA on ranking telescoped datasets (Table \ref{tbl:results-main}), the same embedding model needs to be combined with BM25 to perform well on a raw, unfiltered document collection (Table \ref{tbl:results-main-distractors}).  However, this is not a significant deficiency with the DESM as telescoping is a common initial set in industrial IR pipelines \cite{cambazoglu2010early}.  Moreover, our DESM is especially well suited for late-stage ranking since it incurs little computational overhead, only requiring the document's centroid (which can be precomputed and stored) and its cosine similarity with the query.  

In addition to proposing an effective and efficient ranking scheme, our work suggests multiple avenues for further investigation. Can the IN-IN and the IN-OUT based distances be incorporated into other stages of the IR pipeline, such as in pseudo relevance feedback and for query expansion?  Are there better ways to compose word-level embeddings into document-level representations?  Is there a principled way to filter the noisy comparisons that degrade performance on the non-telescoped datasets?

Content-based document retrieval is a difficult problem.  Not only is language inherently subtle and ambiguous -- allowing for the same ideas to be represented by a multitude of different words -- but the appearance of a given word in a document does not necessarily mean that document is relevant.   While TF-IDF features such as BM25 are a proven source of evidence for \textit{aboutness}, they are not sufficiently precise to rank highly relevant documents ahead of fairly relevant documents.  To do that task well, all of a document's words must be considered.  Neural word embeddings, and specifically our DESM, provide an effective and efficient way for all words in a document to contribute, resulting in ranking attune to semantic subtleties.  

\setlength{\bibsep}{4pt}
\bibliographystyle{abbrvnat}
\raggedright{
\bibliography{desm}
}

\end{document}